\documentclass[letterpaper]{article}
\pdfoutput=1 
\usepackage{jcappub} 
\usepackage{graphicx}
\usepackage{comment}
\usepackage{natbib}

\def\solar{\odot}

\title{
 Accessing the axion via compact object binaries
}

\author[a]{Michael Kavic,}
\author[b]{Steven L. Liebling,}
\author[a]{Matthew Lippert,}
\author[c]{and John H. Simonetti}

\affiliation[a]{Department of Chemistry and Physics, \\SUNY Old Westbury,\\Old Westbury, NY, United States}

\affiliation[b]{Department of Physics, \\Long Island University,\\Brookville, NY, United States}

\affiliation[c]{Department of Physics, \\Virginia Tech, \\Blacksburg, VA, United States}

\emailAdd{kavicm@oldwestbury.edu}
\emailAdd{Steve.Liebling@liu.edu}
\emailAdd{lippertm@oldwestbury.edu}
\emailAdd{jhs@vt.edu}

\abstract{
Black holes in binaries with other compact objects can provide natural venues for indirect detection of axions or other ultralight fields. The superradiant instability associated with a rapidly spinning black hole leads to the creation of an axion cloud which carries energy and angular momentum from the black hole.  This cloud will then decay via gravitational wave emission.  We show that the energy lost as a result of this process tends toward an outspiraling of the binary orbit.  A given binary system is sensitive to a narrow range of axion masses, determined by the mass of the black hole. This proposal provides a complementary alternative to other approaches for detecting or constraining light particles created by superradiance, such as directly measuring the black hole spin or detecting the resulting gravitational wave signal. Pulsar-black hole binaries, once detected in the electromagnetic band, will allow high-precision measurements of black hole mass loss via timing measurements of the companion pulsar. This avenue of investigation is particularly promising in light of the recent preliminary announcements of two candidate black hole-neutron star mergers by LIGO/Virgo (\#S190814bv and \#S190426c). 
We demonstrate that for such a binary system with a typical millisecond pulsar and a 3$M_\odot$ black hole, axions with masses between $2.7 \times 10^{-12}$~eV and $ 3.2 \times 10^{-12}$~eV are detectable.  Recent gravitational wave observations by LIGO/Virgo of binary black hole mergers imply that, for these binaries, gravitational radiation from the rotating quadrupole moment is a dominant effect, causing an inspiraling orbit.  With some reasonable assumptions about the period of the binary when it formed and the spins of the black holes, these observations rule out possible axion masses between $3 \times 10^{-13}$ eV and $6 \times 10^{-13}$ eV.   Future binary black hole observations, for example by LISA, are expected to provide more robust bounds. In some circumstances, neutron stars may also undergo superradiant instabilities, and binary pulsars could be used to exclude axions with certain masses and matter couplings.
}

\begin{document}

\maketitle
\flushbottom

\section{Introduction}
\label{sec:introduction}

Black holes are unique astrophysical laboratories for investigating new physics.  The high curvature near the horizon provides a natural setting in which to explore the physics of strong gravity and high energy.  Until recently, such investigations had been primarily limited to theoretical thought-experiments.  But with the advent of precision, multi-messenger astronomy, black holes may provide observational evidence of new physics.

Through the phenomenon of superradiance,\footnote{For a review, see~\cite{Brito:2015oca}.} black holes can serve as detectors of axions and other extremely light particles.  Any bosonic field with sufficiently small mass can extract energy and angular momentum from rapidly spinning black holes in a manner analogous to the Penrose process \cite{Komissarov:2008yh, Brito:2015oca}.  This superradiant instability eventually results in a coherent, co-rotating boson cloud, carrying a significant fraction of the black hole's original energy \cite{Damour:1976kh,Zouros:1979iw, Detweiler:1980}.  This energy is then slowly dissipated as the cloud decays via boson self-annihilation into gravitational waves \cite{Arvanitaki:2010sy, Yoshino:2013ofa}. The superradiant instability is significant when the boson's Compton wavelength is about the same as the black hole horizon size. For stellar mass black holes, this corresponds to a boson mass of about $10^{-10}$~eV, and for  $10^{10} M_\odot$ supermassive black holes to about $10^{-20}$~eV.

A number of theoretically motivated particles have been proposed which fall within this mass range.  Notably, the QCD axion, initially proposed to solve the strong CP problem, is a pseudo-Goldstone boson whose small mass is generated by QCD instantons and lies at the upper end of this range.  A GUT-scale axion decay constant $f_a = 10^{17}$~GeV corresponds to an axion mass $\mu_a = 6 \times 10^{-11}$~eV. String theory generically predicts a zoo of ultralight axionic particles, termed the ``string axiverse"~\cite{Svrcek:2006yi, Arvanitaki:2009fg, Arvanitaki:2010sy}. These axions result from the Kaluza-Klein reduction of the antisymmetric tensor fields wrapped on one of the many compact cycles.  The masses of these axions are generated by string instantons wrapping those cycles, which yields a range all the way down to $10^{-33}$~eV.

Axions are promising alternatives to WIMPS as dark matter candidates \cite{Preskill:1982cy, Abbott:1982af, Dine:1982ah, Marsh:2015xka}. In order to have the correct abundance, they must be produced non-thermally, for example, via the misalignment mechanism.  Particularly light axions ($\mu_a \lesssim 10^{-22}$~eV) behave as fuzzy dark matter, smoothing out small scale density fluctuations \cite{Hu:2000ke}.  In addition, due to their weakly broken shift symmetry, axions have been employed extensively to construct models of large-field inflation, such as natural inflation and axion monodromy inflation \cite{Freese:1990rb, Silverstein:2008sg}.\footnote{See \cite{Pajer:2013fsa} for a review of axion inflation.} Other possible light particles include, for example, dilatons, dark photons, massive gravitons, or other hidden sector massive vectors and tensors. In this paper we focus on axions, but the discussion and proposed observation applies as well to any ultralight, weakly self-interacting boson.

The advent of gravitational wave astronomy has created the possibility of directly observing the gravitational wave emission from a black hole-axion system.  The decaying axion cloud continuously emits gravitational waves at a single frequency proportional to the axion mass.  For axion clouds with masses in the range $2\times 10^{-13} \lesssim \mu_a \lesssim 10^{-12}$~eV forming around stellar-mass black holes, LIGO may have the potential to detect this continuous signal. LISA, when operational, may be better able to observe continuous emission from clouds of axions with smaller masses, $5\times 10^{-19} \lesssim \mu_a \lesssim 5\times 10^{-16}$~eV, formed around supermassive black holes \cite{Brito:2017wnc, Brito:2017zvb}.

In addition, other observational signatures have been considered. If the axion self-interaction is sufficiently large, the superradiant cloud can rapidly collapse, in a process known as a  ``bosenova," leading to a gravitational wave burst~\cite{Arvanitaki:2014wva}.  For black holes in binary systems, the dynamics of the inspiral results in much richer dynamics, involving resonances which potentially lead to enhanced gravitational wave signals~\cite{Baumann:2018vus, Berti:2019wnn, Baumann:2019ztm}.  An alternative approach is to indirectly infer the existence of the axion cloud. For example, for a given axion mass, black holes in a certain range of corresponding masses and sufficiently high spins will rapidly spin down due to the superradiant instability.  As a result, black holes with those masses and spins will tend not to be observed, leaving underpopulated regions in the mass-spin Regge plane \cite{Arvanitaki:2009fg, Arvanitaki:2010sy}. Another possibility is to analyze the gravitational wave signal of black hole-neutron star~(BH-NS) or neutron star-neutron star~(NS-NS) binary mergers for deviations due to the presence of axions \cite{Hook:2017psm, Sagunski:2017nzb, Huang:2018pbu, Edwards:2019tzf}.

We propose a complementary approach in which the black hole energy transferred to the axion cloud and then dissipated as gravitational waves is measured by its orbital effect on a binary companion.  For a black hole in a binary system, superradiant mass loss pushes the binary towards outspiral, and so this effect is opposite to the usual inspiral resulting from gravitational wave emission due to the binary's orbital motion.  The balancing of these two competing effects appears similarly in the floating orbits of \cite{Cardoso:2011xi}.\footnote{Thanks to Vitor Cardoso for pointing this out.}

An extremely sensitive probe of black hole mass loss is radio emission from a pulsar orbiting the black hole.  For a black hole in a binary with a pulsar, changes in the mass can be precisely measured using pulsar timing.\footnote{The idea of using PSR-BH binaries to test superradiance is not entirely new. In \cite{Rosa:2015hoa}, it was suggested that electromagnetic or gravitational radiation emitted by a pulsar could undergo superradiant scattering off a companion black hole.}  Pulsar-black hole (PSR-BH) binaries have been termed the ``holy grail of astrophysics"~\cite{FaucherGiguere:2010bq} both because of their unique potential as precision astrophysical laboratories but also because no such systems have so far been observed.  However, many pulsar-neutron star~(PSR-NS) binaries have been discovered.  In these systems, pulsar timing allows extraordinarily accurate measurements of the orbital parameters, including the masses and spins of both objects. Most famously, observations of the Hulse-Taylor binary pulsar~(PSR 1913$+$16) were the first to confirm the existence of gravitational waves, decades before their direct detection by LIGO~\cite{1982ApJ...253..908T}.

The prospects for discovery of PSR-BH binaries are promising. The recent preliminary announcements of two candidate BH-NS mergers by LIGO/Virgo (\#S190814bv and \#S190426c) strongly indicate such systems exist \cite{S190814bv,2019GCN.24168....1L,Lattimer:2019qdc}. Many BH-NS binaries are predicted to exist near the galactic center and should be readily detectable by LISA~\cite{AmaroSeoane:2012je}. Such systems could also be discovered through radio observations of pulsar emission. The Square Kilometer Array~(SKA) radio telescope is expected to be able to detect all galactic pulsars~\cite{FaucherGiguere:2010bq}. As we discuss below, multi-messenger observations of such binaries offer a powerful tool to search for axions.

We argue that for each PSR-BH binary discovered, timing observations would be able to detect axions in a narrow mass range, inversely proportional to the black hole mass.  For example, a pulsar-black hole binary with a $3 M_\odot$ black hole\footnote{We consider a $3 M_\odot$ black hole throughout as a fiducial value, as in \cite{Baumann:2018vus}, but the effects are not especially sensitive to this particular value. Also, it should be noted that the limit set in current work for $3 M_\odot$ black holes overlaps with the limit set in \cite{Arvanitaki:2014wva}. However, that limit applies only to the QCD axion. The limits outlined here apply to any axion-like particle, including axions predicted by the string axiverse which, unlike the QCD axion, are generically expected to have ultra-light masses.} would be sensitive to axions with masses between $2.7 \times 10^{-12}$~eV to $ 3.2 \times 10^{-12}$~eV.  For an intermediate mass ($10^4 M_\odot$) black hole, axions in the range  $9 \times 10^{-16}$~eV to $1.3 \times 10^{-14}$~eV would be detectable.


One caveat is that the black hole must have been spinning sufficiently rapidly in the past to trigger the formation of an axion cloud.  Pulsar timing observations may be able to measure the black hole spin with reasonable accuracy, at least for close orbits and high spins \cite{Liu:2014uka}.  The observation of a PSR-BH binary with sufficiently high black hole spin 
would constitute a null detection and would rule out axions within the corresponding mass range \cite{Arvanitaki:2009fg, Arvanitaki:2010sy}. 
However, if a binary with a low-spin black hole is observed, this could be the result of a black hole having already rapidly spun down via superradiance and formed an axion cloud.  In this case, accurate measurements of the orbital period would detect the mass loss due to the slow decay of the cloud.

In addition to PSR-BH binaries, binary black holes (BBH), whose mergers are currently being observed by LIGO/Virgo, are another system in which to probe superradiant mass loss.  Such mergers are the end result of binary inspiral, so we can infer that the superradiant mass loss which leads to outspiraling is necessarily limited.  Together with the measurement of the black holes' pre-merger spins, BBH mergers can be used to exclude a range of axion masses.  Consider, for example, the first observed binary black hole merger GW150914.  Assuming that the binary inspiralled from a period of at least 57 hours and at least one of the black holes had an initial spin of 50\% of maximal, axion masses from $3\times 10^{-13}$~eV to $6\times 10^{-13}$~eV would be excluded.

In addition to the many BBH mergers observed via current detectors,
LISA and third-generation ground-based gravitational wave detectors are expected to detect quasi-stable BBH systems whose inspiral or outspiral would be observed. These, along with improved black hole spin measurements, will allow a larger and more robust probe of the axion mass range.

Neutron stars may also be subject to superradiant instabilities, populating an axion cloud which then decays via gravitational radiation \cite{Cardoso:2015zqa, Day:2019bbh, Kaplan:2019ako}. The expected range of neutron star masses is rather limited, from about $1.4 M_\odot$ to roughly $2.5 M_\odot$, so this effect would be sensitive only to a narrow range of axion masses.  However, unlike the black hole case, neutron star superradiance is not universal and depends on details of axion-matter couplings.

In Sec.~\ref{sec:superradiance}, we discuss black hole superradiance in more detail, in particular the formation of the axion cloud and emission of gravitational waves.  In addition, we discuss enhanced gravitational wave emission due to binary resonance effects, as well as the possibility of neutron star superradiance.  Sec.~\ref{sec:measuring} presents the proposed astronomical observations and gives estimates of the ranges of axion masses for which the black hole's mass loss would be observable.  We begin in Sec.~\ref{sec:PSR-BH} with timing measurements from PSR-BH binaries, followed in Sec.~\ref{sec:BH-BH} by gravitational wave observations of binary black hole systems.  We discuss possible timing measurements of PSR-NS binaries in Sec.~\ref{sec:PSR-NS}. Finally, in Sec.~\ref{sec:discussion} we conclude with discussion and open questions.

\section{Black hole superradiance and spin down}
\label{sec:superradiance}

Black hole superradiance is a classical effect in which a wave incident upon a spinning black hole is amplified, in the process extracting energy and angular momentum from the black hole. If the angular phase velocity of the wave is lower than the angular velocity of the black hole, the wave receives a kick from the more rapidly spinning black hole.\footnote{Black hole superradiance is in many ways analogous to a wave version of the Penrose process of black hole energy extraction.  However, superradiance requires a horizon while the Penrose process relies only on an ergoregion \cite{Brito:2015oca}.}

This type of rotational superradiance is a general phenomenon and can occur in any system with a rotating, dissipating surface.  In the case of black holes, this surface is the event horizon. Superradiance can also take place, for example, near spinning neutron stars as a result of dissipative interactions with the magnetosphere of the neutron star \cite{Cardoso:2015zqa, Day:2019bbh}.

Ultralight particles, such as axions, can be subject to superradiance if their Compton wavelength $\lambda_a = \hbar/\mu_a$ is long compared to the size of the black hole which, for a black hole of mass $M_\mathrm{BH}$, is set by the length scale $R_\mathrm{BH}=GM_\mathrm{BH}$.  In this case, on the black hole length scale, the axion is wavelike or ``fuzzy." Compared with a stellar-mass black hole with $M_\mathrm{BH} = 3 M_\odot$, for example, an axion with $\mu_a \lesssim 10^{-11}$~eV would be wavelike.

The states of a fuzzy axion near a black hole are analogous to the states of an electron in a hydrogen atom, with gravity rather than electromagnetism providing the binding force.  For this black hole atom, the coupling constant analogous to the fine structure constant is the ratio of these two length scales
\begin{equation}
    \alpha \equiv \frac{R_\mathrm{BH}}{\lambda_a} = \frac{GM_\mathrm{BH} \mu_a}{\hbar} = \left(\frac{M_\mathrm{BH}}{3M_\solar}\right) \left(\frac{\mu_a}{4.5\times 10^{-11}  \mathrm{eV}}\right)\ .
\end{equation}
The orbitals of the axion are those of the hydrogen atom, labeled by the principal $n$, orbital $l$, and magnetic $m$ quantum numbers, with the energy levels given by
\begin{equation}
\label{eq:energylevels}
    \omega_n = \mu_a - \frac{\mu_a \alpha^2}{2 n^2} + \mathcal{O}(\alpha^4) \ .
\end{equation}

However, the black hole-axion system is different in two important respects.  First, axions are bosons, and so a given state can be populated by an arbitrary number of axions.  And, due to the black hole horizon, these states are unstable.

\subsection{Superradiant instability and axion cloud formation}
\label{sec:cloudformation}

Axion modes with angular velocity less than that of the black hole will be subject to superradiance.  For a black hole with mass $M_\mathrm{BH}$, angular momentum $J_\mathrm{BH} = a M_\mathrm{BH}$, and horizon radius $R_+ = G\left(M_\mathrm{BH} + \sqrt{M_\mathrm{BH}^2 - a^2}\right)$, the angular velocity is 
\begin{equation}
    \Omega_\mathrm{BH} = \frac{a}{2 M_\mathrm{BH} R_+} \ .
\end{equation}
The condition for superradiance is then
\begin{equation}
\label{eq:superradiantcondition}
   \frac{\omega_n}{m} <  \frac{a}{2 M_\mathrm{BH} R_+}\ .
\end{equation}
Approximating Eq.~\eqref{eq:energylevels}  as $\omega_n \approx \mu_a$ and noting that $a/R_+ \leq 1$ allows the superradiance condition \eqref{eq:superradiantcondition} to be written as
\begin{equation}
\label{eq:maxalpha}
   \alpha \lesssim  \frac{m}{2} \frac{a}{R_+} < \frac{m}{2} \ .
\end{equation}

For massive fields, modes amplified by superradiance are reflected off the effective potential and re-scattered off the black hole.  This repeated amplification leads to an instability with exponentially growing occupation numbers for modes satisfying \eqref{eq:superradiantcondition}, while modes not satisfying \eqref{eq:superradiantcondition} decay. 

The fastest growing mode is $\{n,l,m\} = \{2,1,1\}$, and the growth rate $\Gamma$ increases rapidly with $\alpha$, up to a maximum $\Gamma_{max}$ in the range $0.1 \lesssim \alpha \lesssim 0.5$ \cite{Dolan:2007mj, Arvanitaki:2010sy}. For $\alpha \ll 1$, $\Gamma$ can be approximated as \cite{Detweiler:1980}
\begin{equation}
\label{eq:Gammasmallalpha}
   \Gamma = \frac{1}{24} \frac{a}{M_\mathrm{BH}} \alpha^9 \ .
\end{equation}
The maximum growth rate $\Gamma_{max}$ increases with black hole spin $a$, as does the $\alpha$ for which $\Gamma$ is maximized.  For $a/M_\mathrm{BH} = 0.7$, $\Gamma_\mathrm{max} = 3.3 \times 10^{-10}\, R_\mathrm{BH}^{-1}$ which occurs at $\alpha_\mathrm{max} = 0.187$, while for $a/M_\mathrm{BH} = 0.99 $, $\Gamma_\mathrm{max} = 1.5 \times 10^{-7}\, R_\mathrm{BH}^{-1}$ at $\alpha_\mathrm{max} = 0.42$
(see Table~1 of~\cite{Dolan:2007mj}). 

For $\alpha > \alpha_\mathrm{max}$, $\Gamma$ falls rapidly, and the instability in the $m=1$ mode disappears when $\alpha$ reaches the superradiance bound \eqref{eq:maxalpha}.  For $a/M_\mathrm{BH}=0.5 $, the $m=1$ mode become stable at $\alpha \approx 0.13$, and for $a/M_\mathrm{BH}=0.999$, the limiting value is $\alpha \approx 0.5$  (see Fig.~6 of \cite{Dolan:2007mj}).

For larger values of $\alpha$, the leading instability shifts to modes with larger values of $m$.  For a given $\alpha$, there will always be modes with large enough $m$ to satisfy \eqref{eq:maxalpha}.  However, the rate at which these modes become populated falls exponentially with $\alpha$.  In the regime $\alpha \gg 1$, $\Gamma$ can be approximated as \cite{Zouros:1979iw, Arvanitaki:2010sy}
\begin{equation}
\label{eq:Gammalargealpha}
   \Gamma \approx  10^{-7} e^{-3.7\alpha} \, R_\mathrm{BH}^{-1} \ ,
\end{equation}

The instability analysis of \cite{Dolan:2007mj, Arvanitaki:2010sy} indicates the formation of a classical axion condensate.  The end result is an axion cloud of size $R_c \sim \frac{n^2}{\alpha^2} R_\mathrm{BH} > R_\mathrm{BH}$ orbiting the black hole. Numerical simulations including nonlinear back reaction, for example \cite{East:2017ovw}, support this conclusion.

The formation time of this axion cloud $\tau_c$ can be estimated from the growth rate: $\tau_c \approx 1/\Gamma$.   For $\alpha \ll 1$ or $\alpha \gg 1$, superradiant growth is so slow that no appreciable axion cloud forms in the lifetime of the binary.  However, if $\alpha \approx \alpha_\mathrm{max}$, the cloud forms comparatively rapidly.  For example, for a black hole with $a = 0.99 M_\mathrm{BH}$, if $\alpha \approx  0.42$, the formation time would be
\begin{equation}
\label{eq:tauc}
   \tau_c \gtrsim 2\times 10^2 \, {\rm s} \left(\frac{3 M_\solar}{M_\mathrm{BH}}\right) \ .
\end{equation}

There are other factors which affect the relevant timescale for the instability. For example, vector or tensor fields can be unstable at higher rates and shorter timescales \cite{Pani:2012bp}. Furthermore, the above estimates are for isolated black holes.  As discussed in Sec.~\ref{sec:resonances}, for black holes in a binary system, other effects, such as resonances, can significantly alter the dynamics. 

\subsection{Axion cloud depletion and gravitational wave emission}
\label{sec:GWemission}

Once the rotating axion cloud forms, it begins to decay by emitting gravitational waves.\footnote{This assumes a real axion field, whose clouds are necessarily time-dependent and non-axisymmetric.  Complex fields, in contrast, can form  time-independent, axisymmetric configurations which inhibit this type of gravitational radiation.}  Pairs of axions can annihilate into a single graviton whose frequency is  $2\omega_n \approx 2 \mu_a$, resulting in a continuous, monochromatic signal.\footnote{Gravitational radiation can also result from level transitions, in analogy with photon emission in an atom \cite{Arvanitaki:2010sy}.  However, we will assume the axion cloud is formed primarily by condensation of the leading unstable mode $\{2,1,1\}$, in which case this mechanism is subdominant.} 

The strength of gravitational wave emission is strongly dependent on $\alpha$.  For an axion cloud of mass $M_c$, the power emitted is \cite{Yoshino:2013ofa, Brito:2017zvb, Baumann:2018vus} 
 \begin{equation}
 \label{eq:GWpower}
 \dot E_\mathrm{GW} \approx 0.01 \left(\frac{M_c}{M_\mathrm{BH}}\right)^2\alpha^{14}  \ ,
 \end{equation}
where the approximation is accurate for $\alpha \lesssim 0.1$.  For larger values of $\alpha$, however, \eqref{eq:GWpower} is an overestimate due to nonlinear effects.  

Assuming this gravitational wave emission is the dominant channel by which the axion cloud decays, then $\dot E_\mathrm{GW} = - \dot M_c$.  The mass of the cloud as a function of time can then be found by solving \eqref{eq:GWpower} to give
 \begin{equation}
 \label{eq:GWpower2}
 M_c(t) \approx  \frac{M_{c0}}{1 + \frac{t}{\tau_d}} \ ,
 \end{equation}
where $M_{c0}$ is the initial mass of the axion cloud at $t=0$ and $\tau_d$ is the decay time of the cloud.  For a black hole whose rotation is near maximal, the initial mass of the axion cloud can be approximated as $M_{c0} \approx \alpha M_\mathrm{BH}$~\cite{Brito:2017zvb, Baumann:2018vus}.\footnote{This approximation is valid for  $\alpha \ll 1$.  For larger values of $\alpha$, numerical simulations such as \cite{East:2017ovw} are needed.}  

From \eqref{eq:GWpower2}, the rate at which the axion cloud loses mass is then approximately
 \begin{equation}
     \dot M_c \approx -2 \times 10^{-8} \frac{M_\solar}{ \mathrm{yr}} \left(\frac{\alpha}{0.07}\right)^{16} 
     \label{eq:massloss}
 \end{equation}
over a time scale set by the decay time of the cloud,
\begin{equation}
\label{eq:taud}
\tau_d \approx 10^7\, \mathrm{yr} \left( \frac{M_\mathrm{BH}}{3 M_\odot} \right) \left( \frac{0.07}{\alpha} \right)^{15}
\approx 10^9\, \mathrm{yr} \left( \frac{M_\mathrm{BH}}{10^5 M_\odot} \right) \left( \frac{0.1}{\alpha} \right)^{15} \ .
\end{equation}

The time $\tau_c$ for the axion cloud to form via superradiance \eqref{eq:tauc} is typically much shorter than the depletion time $\tau_d$ via gravitational waves \eqref{eq:taud}, so it is reasonable to treat the processes of formation and dissipation as occurring sequentially.

The very strong dependence of $\dot M_c$ and $\tau_d$ on $\alpha$ implies a narrow range for which either the gravitational waves or the resulting mass loss would be observable.  For $\alpha \ll 0.1$, the decay process lasts a long time, but the mass is emitted very gradually by very weak gravitational waves.  On the other hand, for $\alpha > 0.1$, the cloud decays very rapidly such that any black hole observed would mostly likely have already completed this process.

For a stellar-mass black hole, the optimal value for observation is around $\alpha \approx 0.07$. In this case, the axion cloud decay will be dynamically relevant on the same time scale as the lifetime for a typical compact binary system, which is on the order of $10^7\, \mathrm{yr}$.  For a binary involving a supermassive black hole, a longer cloud lifetime implies a somewhat larger range of $\alpha$ for which the decay process could be observable.

\subsection{Binaries and resonances}
\label{sec:resonances}

In Secs.~\ref{sec:cloudformation} and~\ref{sec:GWemission}, we have focused on the process of black hole superradiance and gravitational wave emission for an isolated black hole, without reference to a binary companion.  However, since our proposal involves observing black hole mass loss through the orbital effect on a binary companion, we need to ask what effect that companion has on the superradiance process.

It has been recently shown~\cite{Baumann:2018vus, Berti:2019wnn, Baumann:2019ztm} that the presence of a binary companion can complicate the dynamics of the cloud, mixing axion modes and leading to much richer predictions for observable gravitational wave emission.  In particular, certain binary orbits can trigger resonant transitions, which can significantly accelerate the depletion of the cloud and produce a greatly enhanced, but short-lived gravitational wave signal.  As discussed in Sec.~\ref{sec:observingresonances} below, the associated transient enhanced mass loss rates would lead to more rapid and observable outspiral of the companion's orbit.  However, we will leave a thorough analysis of these effects for future work.

Another possible set of effects are due to the companion passing through the axion cloud.  Dynamical friction would be expected to exert a drag force on the companion, enhancing the rate of binary inspiral~\cite{Edwards:2019tzf,Zhang:2019eid}. Alternatively, the orbital motion of the companion could excite superradiant modes and extract angular momentum from the cloud, counteracting the inspiral and leading to floating orbits~\cite{Cardoso:2011xi}. In addition, for a PSR-BH binary, the electromagnetic pulsar signal may be modified due to axion-photon couplings between the axion cloud and the neutron star plasma~\cite{Edwards:2019tzf}.

However, because the size of the cloud is not that much larger than the horizon size of the black hole, these effects would only be expected to occur very late in the inspiral process, close to the merger, when the orbital radius is very small.  Our focus will instead be on larger, more stable binaries, for which the companion's orbit is well outside the axion cloud.  In this case,  the companion only probes the superradiant effects gravitationally.

Furthermore, the binary itself could also be subject to a superradiant instability~\cite{Wong:2019kru}. In this case, the orbital rotation of the binary would trigger the superradiant formation of an axion cloud.  The relevant length scale would be the orbital radius of the binary, rather than just the radius of the black hole, meaning the superradiant effect would appear at a correspondingly lower axion mass scale.  The resulting formation of an axion cloud would spin down the binary, leading to an enhanced rate of inspiral.  This is very similar to the way in which the related Blandford-Znajek effect, by which a rotating black hole spins down in the presence of a magnetic field, has also been shown to occur in black hole binaries~\cite{Neilsen:2010ax}.  In the case of a single black hole, superradiance relies on the spin being sufficiently high, but this can be difficult to measure.  In contrast, for a black hole binary, the binary's orbital motion can be accurately determined from the gravitational wave signal.

\subsection{Neutron star superradiance}
\label{sec:NSsuperradiance}

Superradiance is a general phenomenon which can occur in rotating systems provided there is a mechanism for dissipation.  For black holes, the horizon provides a dissipative surface, and since all matter couples to gravity, black hole superradiance is universal. Superradiant instabilities have also been proposed for other compact objects, such as conducting stars~\cite{Cardoso:2015zqa}.  In such cases the necessary dissipation relies on the existence of certain axion-matter couplings.

Recently, it has recently been suggested that rapidly rotating neutron stars may be subject to superradiant instabilities stemming from interactions between the axions and photons in the magnetosphere \cite{Day:2019bbh}.\footnote{Other mechanisms have been proposed by which compact objects could emit axions, which also rely on various axion-matter couplings. For example, NS binaries might emit axionic radiation due to their orbital motion \cite{Poddar:2019zoe, Poddar:2019wvu}.}  The effect depends both on the axion mass and the size of the axion-photon coupling.  However, once the axion cloud is formed, the decay by gravitational wave emission is expected to proceed as in the black hole case.

Although neutron star superradiance is still relatively speculative and relies on additional assumptions, such an effect is  testable by observing the mass lost via the decay of the axion cloud. 

\section{Searching for axions in binary systems}
\label{sec:measuring}

We consider the observable effect of the superradiant instability discussed above in the case of the three possible binary systems PSR-BH, BH-BH and PSR-NS. We  outline the limits such observations could set on the axion mass.

\subsection{Pulsar-black hole binary timing}
\label{sec:PSR-BH}



First, we consider the possibility of measuring  superradiance-induced mass loss in PSR-BH systems in which the black hole populates the cloud which then proceeds to deplete itself.One might hope to measure the spin of the black hole directly as a function of time, a prospect discussed for general binaries in detail in Ref.~\cite{Liu:2014uka}. However, the measurement is difficult and generally requires the system to be in a close orbit and have a high spin. Depending on the parameters of the system, the precision of a black hole spin measurement can reach somewhere between 1\%-10\%.  More significantly, the superradiant instability has a short time scale, and so, if it occurs, the black hole spins down rapidly. Although the measurement of a high spin would represent a null detection and could exclude a range of axion masses, a low-spin black hole could be the result of an earlier superradiant spin down or simply the black hole was just slowly spinning all along.

Instead, an easier measurement, and hence one which maintains high precision even when the constituents are well separated, is the binary period, $P_b$. Such
a measurement can, in principle, be made with a precision of about $10^{-15}\ {\rm s/s}$, as evidenced by observations of the Hulse-Taylor binary pulsar~\cite{Weisberg:2010zz}.  This measurement can reveal the mass loss associated with the depletion of the axion cloud. 

Consider a black hole which has populated a surrounding axion cloud. As discussed in Sec.~\ref{sec:GWemission}, this cloud proceeds to decay via gravitational radiation, with the black hole-axion system losing mass at a rate given by Eq.~\eqref{eq:massloss}.  To observe this mass loss we turn now to an argument first presented in Ref.~\cite{Simonetti:2010mk}, which studied a very different mass loss mechanism but remains relevant here. As the black hole-axion system slowly loses mass, the semi-major axis of the binary $a$ grows as
\begin{equation}
\label{eq:orbitaldecay}
    \dot a = -\frac{a \dot M_\mathrm{BH}}{M_\mathrm{BH} + M_p} \,
\end{equation}
where $M_p$ is the mass of the pulsar companion. From Kepler's third law $\dot P_b = 2 \dot{a} P_b/a $.
Combining this with \eqref{eq:orbitaldecay} and \eqref{eq:massloss}, choosing $M_\mathrm{BH}=3M_\solar$ and $M_p=1.4M_\solar$, and using a fiducial value for the binary period of 10 hours, one obtains
\begin{equation}
    \dot P_b \approx 10^{-11}\ \mathrm{s/s}\ 
           \left( \frac{M_\mathrm{BH}+M_P}{4.4 M_\solar}\right)^{-1}
           \left( \frac{P_b}{10^{\rm h}} \right)
           \left( \frac{\alpha}{0.07} \right)^{16}.
\label{eq:pbdot}
\end{equation}
To get a sense of how large this is, it may help to compare this value to the expected rate of change of the period due to gravitational wave emission of the binary from general relativity~(GR),
\begin{equation}
    \dot P_{b,\mathrm{GR}} = -2.4\times10^{-13}\ {\rm s/s} \left(\frac{M_\mathrm{BH}}{3M_\odot}\right)
     \left(\frac{M_P}{1.4M_\odot}\right)
     \left(\frac{M_\mathrm{BH}+M_P}{4.4M_\odot}\right)^{-1/3}
     \left( \frac{P_b}{10^{\rm h}} \right)^{-5/3}
     f(e),
\end{equation}
where 
\begin{equation}
    f(e)=\left(1+\frac{73}{24}e^2+\frac{37}{96}e^4\right)
    (1-e^2)^{-7/2}
\end{equation}
is an expression dependent on the eccentricity $e$ of the binary \cite{Simonetti:2010mk}.

More importantly, the value of $\dot P_b \approx 10^{-11}~\mathrm{s/s}$ is about $10^4$ times larger than the $10^{-15}$~s/s precision with which $\dot P_b$ has been measured for the Hulse-Taylor binary. Since this precision allows us to measure values of $\dot P_b$ that are 4 orders of magnitude smaller than given in Eq.~\eqref{eq:pbdot}, we are sensitive to values of $\alpha$ down to about 0.06 for this binary system, or an axion mass lower limit of $2.7\times10^{-12}$~eV.

Of course, larger values of $\alpha$ give a much larger values of $\dot P_b$, but as pointed out in Ref.~\cite{Baumann:2018vus}, given the sensitivity of Eq.~\eqref{eq:massloss} to $\alpha$, values greater than $\alpha=0.07$ will result in a very much shorter lifetime for the axion cloud.  In this case, the mass loss would be so transient that the chance of observing it would be very low. Thus, the practical maximum for observing a binary system with a stellar-mass black hole using this method is $\alpha=0.07$, corresponding to an upper axion mass limit of $3.2\times10^{-12}$~eV. 

Another interesting aspect of this calculation is that the change in period from mass loss by the black hole is opposite that of gravitational wave emission from the binary motion. In other words, if the black hole loses mass, the binary becomes less bound and slowly outspirals.  However, by emission of gravitational radiation the binary system becomes more bound, causing it to inspiral. One can equate the respective expressions for $\dot P_b$ to achieve a critical balance of the two effects. Doing so provides a value of the period, $P_{b,\mathrm{critical}}$, at which the binary is balanced between inspiral and outspiral:
\begin{equation}
     P_{b,\mathrm{critical}} \approx 2.4^{\rm h} \left(\frac{M_\mathrm{BH}}{3M_\odot}\right)^{3/8}
     \left(\frac{M_P}{1.4M_\odot}\right)^{3/8}
     \left(\frac{M_\mathrm{BH}+M_P}{4.4M_\odot}\right)^{1/4}
     f(e)^{3/8} \left(\frac{\alpha}{0.07}\right)^{-6}.
     \label{eq:pcrit}
\end{equation}
Therefore, if a binary with a black hole (PSR-BH or BH-BH) were observed to outspiral, either via a pulsar signal or gravitational waves, then mass loss from the black hole, possibly from superradiance, would be a natural explanation.\footnote{Note, that such an observation of superradiant-induced mass loss is likely degenerate with other  scenarios by which black holes might lose mass, such as the existence of extra dimensions~\cite{Simonetti:2010mk}.}

Although the axion mass range for any particular PSR-BH binary is relatively narrow, a population of such systems with a variety of black hole masses would allow a wider range of axion masses to be probed.  We have shown that stellar-mass binaries are sensitive to a  rather small range of values of $\alpha$ from 0.06 to 0.07.
We could instead consider binaries with intermediate mass black holes up to roughly $10^4 M_\odot$. If we assume the same pulsar timing precision and orbit determination, such a binary would be sensitive to axion masses at a lower bound of about $10^{-15}$~eV. However, such black holes may be located in crowded stellar neighborhoods, perhaps at the center of globular clusters. If, on the other hand, one restricts the use of the pulsar-timing technique to stellar-mass black holes only (avoiding the necessity of analyzing systems in crowded stellar neighborhoods), then an upper mass limit of about $10^2 M_\odot$ would be reasonable, and the lower bound on the range of axion masses accessible would be about $10^{-13}$~eV. 
Thus the full range of black hole masses accessible to this observational technique is 3 to $10^4M_\odot$, corresponding to axion masses of $3\times10^{-12}$ to $10^{-15}$~eV. At the lower black hole-mass end of this range, the practical values of $\alpha$ range from about 0.06 to 0.07, while at the upper end of the black hole-mass range, the $\alpha$ range is about 0.07 to 0.1.

\subsubsection{Observing Binary Resonances in a PSR-BH system}
\label{sec:observingresonances}
As noted above in Sec.~\ref{sec:resonances}, binary systems in which one of the constituents has generated an axion cloud can experience resonance events during which the cloud can be rapidly depleted. The conditions under which a binary undergoes a resonant phase in its evolution are discussed in detail in \cite{Baumann:2018vus, Baumann:2019ztm}. Depending on the parameters of the system, in particular the mass ratio of the pair, the effect of such resonances can be quite significant and the depletion of the cloud can be extremely fast. In contrast to the slower cloud depletion discussed in Sec.~\ref{sec:GWemission}, which neglects the presence of the companion, the rapid depletion due to resonances is likely to overwhelm the tendency toward inspiral due to gravitational wave emission.

Given that these resonances can lead to such dramatic effects, it makes sense to consider where in the binary evolution they take place. The resonant orbital period can be computed from Eq.~(4.14) of  Ref.~\cite{Baumann:2018vus}. We see that, for a binary with a 3$M_\odot$ black hole and for $\alpha=0.1$, resonances are expected to appear at orbital periods of roughly 300 and 3 seconds. These periods are much smaller than the fiducial period we consider here of 10 hours and occur close to merger.  We will therefore defer a more thorough investigation of resonant effects to future work.

\subsubsection{Multi-messenger observations of PSR-BH binaries}
\label{sec:multimessenger}

Our discussion has so far focused on radio observations of a PSR-BH system. This, of course, presupposes the detection of such a system. We wish to consider the observational prospects for such a system both in terms of initial detection and sustained multi-messenger observations. 

The Square Kilometer Array~(SKA) radio telescope is expected to be able to discover all the pulsars in the galaxy that beam radiation toward the Earth, including pulsars in binary systems with black holes~\cite{Levin:2017mkq}. It will also be a premier instrument for pulsar timing observations. While the expected number of such systems is uncertain, the increased density of compact objects near the galactic center may increase the likelihood of such a system being found once this region can effectively be searched for pulsars.
The observation proposed above in Sec.~\ref{sec:PSR-BH} relies on the high precision with which pulsar timing can measure the change in period of a PSR-BH binary. Of course, such a system is also a prime candidate source for gravitational wave observatories. 

A gravitational wave signal accompanying the pulsar timing can, in principle, provide valuable additional information. 
In particular, gravitational wave observations could help measure the black hole spin.  Although current gravitational wave detectors are not terribly sensitive to the spin generally, future generations of  detectors are expected to greatly improve such measurements.  Since there are other mechanisms by which black holes might lose mass, for example \cite{Simonetti:2010mk}, gravitational wave observations could help break this degeneracy. 

Such a dual band observation presents timing problems, however. Current ground-based gravitational wave detectors are only sensitive to a relatively short window at the end of the inspiral. 
The strong gravitational wave emission which facilitates  detection also results in a rapid inspiral, which would overwhelm any outspiraling due to superradiant mass loss. Furthermore, a merging binary is so late in its evolution that, if there were a superradiant instability, the black hole would most likely have long since been spun down and the axion cloud depleted.

Instead,  one would ideally hope to catch such a binary long before it merges. In particular, LISA should be sensitive to such a binary. For a galactic PSR-BH in a quasi-circular orbit, one would hope to observe both the (electromagnetic) pulsar timing with the SKA and the gravitational wave emission with LISA. If electromagnetic pulsar timing finds an interesting signal, for example indicating an outspiraling binary, then this could be used for a targeted gravitational wave search that would otherwise be sub-threshold (i.e. below the standard SNR threshold required for detection). Conversely, a detection of a stable NS-BH system by LISA could be followed up by targeted radio observations to search for pulsar emission to try to identify a system from which both electromagnetic and gravitational wave signals could be observed. If the binary separation is not too large, one could observe the binary slowly tighten over perhaps a decade as it enters the LIGO band and merges~\cite{Sesana:2016ljz}. New mid-band detectors in the decihertz region that have been proposed, such as TianGO and DECIGO, might be ideal both for multi-year tracking of compact object binaries and for measuring individual component spins~\cite{Kuns:2019upi,Sato:2017dkf}.

It is also interesting to consider the prospects for a strong electromagnetic counterpart to the gravitational signal produced by the merger of NS-BH binaries. In order to produce such a signal, the neutron star must to be severely disrupted, as opposed to being swallowed intact by the black hole. This is most likely to occur when the neutron star merges with a highly spinning, prograde black hole.  However, such a black hole is precisely the opposite of what is expected from superradiance, in which the black hole is spun down by the populating a surrounding axion cloud.\footnote{Thanks to Francois Foucart for pointing this out.} By the time the binary merges, one might expect the black hole to have been spun down sufficiently so as to no longer be unstable. Indeed, one could turn this observation around to argue that a strong electromagnetic counterpart to a BH-NS merger, depending on the constituent masses, may indicate a long-lived, highly spinning black hole and hence enable a bound on possible axion masses.

\subsection{Constraining axion mass with binary black hole observations}
\label{sec:BH-BH}

While we wait for the first observations of PSR-BH binaries, the majority of LIGO/Virgo observations consist of BBH mergers. Because these events result from inspiral of the binary, we can infer that the outspiraling due to superradiant mass loss is less important than the inspiraling from the usual gravitational wave emission due to the rotating quadrupole moment of the binary.  This observation could then be used to exclude a range of axion masses, subject to two caveats.  First, at least one of the black holes must have been spinning sufficiently rapidly for the superradiant instability to have formed an axion cloud.  Unfortunately, current gravitational wave observations poorly constrain the pre-merger spins of the black holes. Second, the relative impact on the binary orbit of the black hole mass loss and the usual gravitational wave emission depends on the size of the binary orbit.  In particular, the rate of inspiraling due to gravitational radiation accelerates as the orbital period decreases, which implies that this effect was smaller in the past when the period was longer. The resulting constraint on the axion mass depends on the initial binary period, with a larger period giving a larger excluded range. In order to make a useful statement about the axion mass, however, assumptions must be made about the period of the binary when it formed.

As a concrete example, the first observed BBH merger GW150914, subject to some assumptions, excludes axion masses from $3\times 10^{-13}$~eV to $6\times 10^{-13}$~eV.  The existence of an axion with mass in that range would have prevented the merger from taking place.  We assume that the BBH formed in isolation via the collapse of two massive stars already bound together. For a subsequent merger to occur, the initial BBH orbital period must have been less than the critical period given by Eq.~\eqref{eq:pcrit}. Reasonable radii for 35$M_\odot$ and 30$M_\odot$ main sequence stars are $\sim15R_\odot$ \cite{2007hsaa.book.....Z}. For these masses Kepler's third law then yields 57 hours as the least possible value of $P_b$ and thus $P_{b,\mathrm{critical}}$. Then, Eq.~\eqref{eq:pcrit} implies $\alpha\lesssim 0.065$, a result insensitive to which black hole has the axion cloud (or whether both have a axion clouds).\footnote{Because of the high power of $\alpha$ in \eqref{eq:pcrit}, the axion mass limit is  insensitive to the exact values of the black hole masses.} The resulting axion mass limit is $\mu_a \lesssim 3 \times 10^{-13}$~eV.

On the other hand, if the axion mass is too large, the superradiant instability will not occur at all.  The maximum value of $\alpha$ depends on the black hole spin.  Assuming at least one of the black holes is measured to have a pre-merger spin 50\% of maximal, then superradiance would occur for $\alpha \lesssim 0.13$ \cite{Dolan:2007mj}.  The observation of a $30M_\odot$ black hole merger then implies that axion masses must be $\mu_a \gtrsim 6 \times 10^{-13}$~eV.  Note that the upper end of the excluded mass range is dependent on the pre-merger spins of the black holes, which was not well measured by LIGO \cite{LIGOScientific:2018jsj}.  

As discussed in \ref{sec:multimessenger}, future observations with LISA and third-generation ground-based gravitational wave detectors will allow for improved axion searches.  In particular, by observing binaries well before merger, and therefore with much longer period, and with improved measurements of the black hole spin, observations of inspiraling BBH systems will yield stronger and more confident axion mass constraints.

\subsection{Constraining axion mass with binary pulsar systems}
\label{sec:PSR-NS}

Now we consider a third type of compact binary, PSR-NS binaries, as probes of axions.  Unlike PSR-BH binaries, binary pulsars have been discovered and closely studied. As described above in Sec.~\ref{sec:NSsuperradiance}, spinning neutron stars could be subject to superradiance~\cite{Cardoso:2017kgn, Day:2019bbh}.  In contrast to the black hole, for which the horizon which provides universal dissipation, neutron star superradiance depends on the coupling between axions and the neutron star, in order to provide the necessary dissipation mechanism.

In principle it is possible to search for axions through isolated pulsar observations. A fast spinning pulsar could be observed to look for a continual reduction in its rotational period (beyond that expected from magnetic braking) as it populates an axion cloud as discussed in \cite{Cardoso:2017kgn,Day:2019bbh}. However, such a direct observation would require catching the pulsar while it is rapidly being spun down.

If one instead considers the binary system, detection of any outspiral could potentially signal mass loss due to depletion of the cloud, which occurs much more slowly than the cloud formation. In such a binary pulsar system, the angular momentum of the pulsar would have had to be sufficient to trigger a superradiant instability at some point in the past.  Now such a system would feature a pulsar with a relatively stable, albeit sub-instability, rotational period. But because the lifetime of the axion cloud would far exceed the time taken to populate the cloud, the gravitational wave emission would continue causing the binary pair to outspiral or at least to inspiral more slowly than predicted. 

Such an outspiral event would serve as a clear detection of an axion cloud. However, a null detection would not exclude a range of axion masses because it can not be known, in such a system, that the pulsar had previously possessed the required angular momentum to generate an axion cloud.  The only way to rule out certain axion masses would be to observe a sufficiently highly spinning pulsar in a binary which is inspiraling at the usual rate predicted by GR.

Neutron star spins are limited by a theoretical maximum of about $a/M \approx 0.7$, above which the neutron star would break up and fly apart \cite{Cook:1993qr, Cipolletta:2015nga}, which corresponds to a rotational frequency of about 1.4 kHz.  Ideally, one would find a PSR-NS binary with a maximally spinning pulsar and determine whether it outspirals or not. 

However, no near-maximally spinning neutron stars have been found.  The fastest observed pulsar PSR J1748-2446ad has a frequency of 716 Hz \cite{Hessels:2006ze}.  In fact, neutron stars are generally observed to be spinning significantly slower than this maximum. Many explanations have been suggested to explain this gap, including superradiant spindown \cite{Cardoso:2017kgn}. Pulsars in binaries with either a neutron star or a black hole are generally spinning more slowly than the fastest observed pulsars. The Hulse-Taylor pulsar, for example, spins at only 59 Hz \cite{Weisberg:2010zz}.  Not only is this probably below the threshold for superradiance, but such a small fraction of the total energy is rotational kinetic energy that there would not be a measurable amount of mass to lose.


\section{Discussion}
\label{sec:discussion}

We have considered compact object binaries as indirect detectors of axions and other ultralight bosons.  If axions exist within a particular mass range, rapidly rotating black holes (or possibly neutron stars) experience a superradiant instability, populating a surrounding axion cloud which then slowly decays. The dynamics of the binary are determined by two competing effects, a tendency to inspiral from gravitational wave emission versus a tendency to outspiral from the mass loss of the black hole companion.  A stable PSR-BH binary would be ideal for this purpose. Precise electromagnetic measurements of the pulsar should clearly reveal the mass loss associated with the depletion of the axion cloud about the companion black hole.  A black hole in a binary with another black hole, namely a BBH, would still be subject to superradiance,  but the resulting mass loss will generally be unobservable by electromagnetic observations. Instead, we discussed the implications of BBH mergers already observed by LIGO and suggested that observations of BBH systems by LISA will perhaps be ideal for either observing axions or excluding certain axion masses. We also mention that superradiance may occur for rapidly spinning neutron stars, depending on possible couplings of axions to neutron star matter.

Of course, our proposal depends on a number of caveats. The most significant of these is the requirement that the black hole was spinning sufficiently rapidly to undergo superradiance and generate an axion cloud. If indeed this process occurs, the resulting mass loss will detectably counteract the tendency of the binary to inspiral. However, if no mass loss is observed, excluding a range of axion masses requires an accurate measurement of the black hole spin.

The discovery of a stable PSR-BH system would provide a clear testing ground for the existence of axions within a particular mass range. However, such systems have yet to be found. Expected advances in pulsar and gravitational wave detection will make finding such a system considerably more likely in the near future. In broader terms the results presented here demonstrate the potential for compact object binaries to serve as ready-made laboratories to probe the frontier of particle physics.

\acknowledgments
We would like to thank Daniel Baumann, Vitor Cardoso, Francois Foucart, Jonah Kanner, Ken Olum, Paolo Pani, Andrea Possenti, and Frans Pretorius for helpful discussions.
This research is supported by the National Science Foundation grants: PHY-1820733~(ML), PHY-1912769~(SLL), and NSF PHY-1827573~(SLL).
M.~L.~would like to thank the KITP for hospitality and support under the KITP Scholars program, via the National Science Foundation under Grant No. NSF PHY-1748958.

\providecommand{\href}[2]{#2}\begingroup\raggedright
\endgroup


\begin{thebibliography}{10}

\bibitem{Brito:2015oca}
R.~Brito, V.~Cardoso and P.~Pani, \emph{{Superradiance}},
  \href{http://dx.doi.org/10.1007/978-3-319-19000-6}{\emph{Lect. Notes Phys.}
  {\bf 906} (2015) pp.1--237}, [\href{http://arxiv.org/abs/1501.06570}{{\tt
  1501.06570}}].

\bibitem{Komissarov:2008yh}
S.~S. Komissarov, \emph{{Blandford-Znajek mechanism versus Penrose process}},
  \href{http://dx.doi.org/10.3938/jkps.54.2503}{\emph{J. Korean Phys. Soc.}
  {\bf 54} (2009) 2503--2512}, [\href{http://arxiv.org/abs/0804.1912}{{\tt
  0804.1912}}].

\bibitem{Damour:1976kh}
T.~Damour, N.~Deruelle and R.~Ruffini, \emph{{On Quantum Resonances in
  Stationary Geometries}},
  \href{http://dx.doi.org/10.1007/BF02725534}{\emph{Lett. Nuovo Cim.} {\bf 15}
  (1976) 257--262}.

\bibitem{Zouros:1979iw}
T.~J.~M. Zouros and D.~M. Eardley, \emph{{Instabilities of Massive Scalar
  Perturbations of a Rotating Black Hole}},
  \href{http://dx.doi.org/10.1016/0003-4916(79)90237-9}{\emph{Annals Phys.}
  {\bf 118} (1979) 139--155}.

\bibitem{Detweiler:1980}
S.~Detweiler, \emph{Klein-gordon equation and rotating black holes},
  \href{http://dx.doi.org/10.1103/PhysRevD.22.2323}{\emph{Phys. Rev. D} {\bf
  22} (Nov, 1980) 2323--2326}.

\bibitem{Arvanitaki:2010sy}
A.~Arvanitaki and S.~Dubovsky, \emph{{Exploring the String Axiverse with
  Precision Black Hole Physics}},
  \href{http://dx.doi.org/10.1103/PhysRevD.83.044026}{\emph{Phys. Rev.} {\bf
  D83} (2011) 044026}, [\href{http://arxiv.org/abs/1004.3558}{{\tt
  1004.3558}}].

\bibitem{Yoshino:2013ofa}
H.~Yoshino and H.~Kodama, \emph{{Gravitational radiation from an axion cloud
  around a black hole: Superradiant phase}},
  \href{http://dx.doi.org/10.1093/ptep/ptu029}{\emph{PTEP} {\bf 2014} (2014)
  043E02}, [\href{http://arxiv.org/abs/1312.2326}{{\tt 1312.2326}}].

\bibitem{Svrcek:2006yi}
P.~Svrcek and E.~Witten, \emph{{Axions In String Theory}},
  \href{http://dx.doi.org/10.1088/1126-6708/2006/06/051}{\emph{JHEP} {\bf 06}
  (2006) 051}, [\href{http://arxiv.org/abs/hep-th/0605206}{{\tt
  hep-th/0605206}}].

\bibitem{Arvanitaki:2009fg}
A.~Arvanitaki, S.~Dimopoulos, S.~Dubovsky, N.~Kaloper and J.~March-Russell,
  \emph{{String Axiverse}},
  \href{http://dx.doi.org/10.1103/PhysRevD.81.123530}{\emph{Phys. Rev.} {\bf
  D81} (2010) 123530}, [\href{http://arxiv.org/abs/0905.4720}{{\tt
  0905.4720}}].

\bibitem{Preskill:1982cy}
J.~Preskill, M.~B.~Wise and F.~Wilczek,
  \emph{{Cosmology of the Invisible Axion}},
  \href{http://dx.doi.org/10.1016/0370-2693(83)90637-8}{\emph{Phys. Rev.} {\bf
  B120} (1983) 127-132}.


\bibitem{Abbott:1982af}
L.~Abbott and P.~Sikivie,
 \emph{{A Cosmological Bound on the Invisible Axion}},
  \href{http://dx.doi.org/10.1016/0370-2693(83)90638-X}{\emph{Phys. Rev.} {\bf
  B120} (1983) 133-136}.
  

\bibitem{Dine:1982ah}
M.~Dine and W.~Fischler,
 \emph{{The Not So Harmless Axion}},
  \href{http://10.1016/0370-2693(83)90639-1}{\emph{Phys. Rev.} {\bf
  B120} (1983) 137-141}.



\bibitem{Marsh:2015xka}
D.~J.~E. Marsh, \emph{{Axion Cosmology}},
  \href{http://dx.doi.org/10.1016/j.physrep.2016.06.005}{\emph{Phys. Rept.}
  {\bf 643} (2016) 1--79}, [\href{http://arxiv.org/abs/1510.07633}{{\tt
  1510.07633}}].

\bibitem{Hu:2000ke}
W.~Hu, R.~Barkana and A.~Gruzinov, \emph{{Cold and fuzzy dark matter}},
  \href{http://dx.doi.org/10.1103/PhysRevLett.85.1158}{\emph{Phys. Rev. Lett.}
  {\bf 85} (2000) 1158--1161},
  [\href{http://arxiv.org/abs/astro-ph/0003365}{{\tt astro-ph/0003365}}].

\bibitem{Freese:1990rb}
K.~Freese, J.~A. Frieman and A.~V. Olinto, \emph{{Natural inflation with pseudo
  - Nambu-Goldstone bosons}},
  \href{http://dx.doi.org/10.1103/PhysRevLett.65.3233}{\emph{Phys. Rev. Lett.}
  {\bf 65} (1990) 3233--3236}.

\bibitem{Silverstein:2008sg}
E.~Silverstein and A.~Westphal, \emph{{Monodromy in the CMB: Gravity Waves and
  String Inflation}},
  \href{http://dx.doi.org/10.1103/PhysRevD.78.106003}{\emph{Phys. Rev.} {\bf
  D78} (2008) 106003}, [\href{http://arxiv.org/abs/0803.3085}{{\tt
  0803.3085}}].

\bibitem{Pajer:2013fsa}
E.~Pajer and M.~Peloso, \emph{{A review of Axion Inflation in the era of
  Planck}},
  \href{http://dx.doi.org/10.1088/0264-9381/30/21/214002}{\emph{Class. Quant.
  Grav.} {\bf 30} (2013) 214002}, [\href{http://arxiv.org/abs/1305.3557}{{\tt
  1305.3557}}].

\bibitem{Brito:2017wnc}
R.~Brito, S.~Ghosh, E.~Barausse, E.~Berti, V.~Cardoso, I.~Dvorkin et~al.,
  \emph{{Stochastic and resolvable gravitational waves from ultralight
  bosons}}, \href{http://dx.doi.org/10.1103/PhysRevLett.119.131101}{\emph{Phys.
  Rev. Lett.} {\bf 119} (2017) 131101},
  [\href{http://arxiv.org/abs/1706.05097}{{\tt 1706.05097}}].

\bibitem{Brito:2017zvb}
R.~Brito, S.~Ghosh, E.~Barausse, E.~Berti, V.~Cardoso, I.~Dvorkin et~al.,
  \emph{{Gravitational wave searches for ultralight bosons with LIGO and
  LISA}}, \href{http://dx.doi.org/10.1103/PhysRevD.96.064050}{\emph{Phys. Rev.}
  {\bf D96} (2017) 064050}, [\href{http://arxiv.org/abs/1706.06311}{{\tt
  1706.06311}}].

\bibitem{Arvanitaki:2014wva}
A.~Arvanitaki, M.~Baryakhtar and X.~Huang, \emph{{Discovering the QCD Axion
  with Black Holes and Gravitational Waves}},
  \href{http://dx.doi.org/10.1103/PhysRevD.91.084011}{\emph{Phys. Rev.} {\bf
  D91} (2015) 084011}, [\href{http://arxiv.org/abs/1411.2263}{{\tt
  1411.2263}}].

\bibitem{Baumann:2018vus}
D.~Baumann, H.~S. Chia and R.~A. Porto, \emph{{Probing Ultralight Bosons with
  Binary Black Holes}},
  \href{http://dx.doi.org/10.1103/PhysRevD.99.044001}{\emph{Phys. Rev.} {\bf
  D99} (2019) 044001}, [\href{http://arxiv.org/abs/1804.03208}{{\tt
  1804.03208}}].

\bibitem{Berti:2019wnn}
E.~Berti, R.~Brito, C.~F.~B. Macedo, G.~Raposo and J.~L. Rosa,
  \emph{{Ultralight boson cloud depletion in binary systems}},
  \href{http://dx.doi.org/10.1103/PhysRevD.99.104039}{\emph{Phys. Rev.} {\bf
  D99} (2019) 104039}, [\href{http://arxiv.org/abs/1904.03131}{{\tt
  1904.03131}}].

  \bibitem{Baumann:2019ztm}
D.~Baumann, H.~S.~Chia, R.~A.~Porto and J.~Stout,
  \emph{{Gravitational Collider Physics}},
  \href{http://dx.doi.org/10.1103/PhysRevD.101.083019}{\emph{Phys. Rev.} {\bf
  D101} (2020) 083019}, [\href{http://arxiv.org/abs/1912.04932}{{\tt
  1912.04932}}].



\bibitem{Hook:2017psm}
A.~Hook and J.~Huang, \emph{{Probing axions with neutron star inspirals and
  other stellar processes}},
  \href{http://dx.doi.org/10.1007/JHEP06(2018)036}{\emph{JHEP} {\bf 06} (2018)
  036}, [\href{http://arxiv.org/abs/1708.08464}{{\tt 1708.08464}}].

\bibitem{Sagunski:2017nzb}
L.~Sagunski, J.~Zhang, M.~C. Johnson, L.~Lehner, M.~Sakellariadou, S.~L.
  Liebling et~al., \emph{{Neutron star mergers as a probe of modifications of
  general relativity with finite-range scalar forces}},
  \href{http://dx.doi.org/10.1103/PhysRevD.97.064016}{\emph{Phys. Rev.} {\bf
  D97} (2018) 064016}, [\href{http://arxiv.org/abs/1709.06634}{{\tt
  1709.06634}}].

\bibitem{Huang:2018pbu}
J.~Huang, M.~C. Johnson, L.~Sagunski, M.~Sakellariadou and J.~Zhang,
  \emph{{Prospects for axion searches with Advanced LIGO through binary
  mergers}}, \href{http://dx.doi.org/10.1103/PhysRevD.99.063013}{\emph{Phys.
  Rev.} {\bf D99} (2019) 063013}, [\href{http://arxiv.org/abs/1807.02133}{{\tt
  1807.02133}}].

  \bibitem{Edwards:2019tzf}
T.~D.~Edwards, M.~Chianese, B.~J.~Kavanagh, S.~M.~Nissanke and C.~Weniger,
\emph{{A Unique Multi-Messenger Signal of QCD Axion Dark Matter}},
\href{http://dx.doi.org/10.1103/PhysRevLett.124.161101}
{\emph{Phys. Rev. Lett.} {\bf 124} (2020), 161101},
 [\href{http://arxiv.org/abs/1905.04686}{{\tt1905.04686}}].


\bibitem{Cardoso:2011xi}
V.~Cardoso, S.~Chakrabarti, P.~Pani, E.~Berti and L.~Gualtieri, \emph{{Floating
  and sinking: The Imprint of massive scalars around rotating black holes}},
  \href{http://dx.doi.org/10.1103/PhysRevLett.107.241101}{\emph{Phys. Rev.
  Lett.} {\bf 107} (2011) 241101}, [\href{http://arxiv.org/abs/1109.6021}{{\tt
  1109.6021}}].

\bibitem{Rosa:2015hoa}
J.~G. Rosa, \emph{{Testing black hole superradiance with pulsar companions}},
  \href{http://dx.doi.org/10.1016/j.physletb.2015.07.063}{\emph{Phys. Lett.}
  {\bf B749} (2015) 226--230}, [\href{http://arxiv.org/abs/1501.07605}{{\tt
  1501.07605}}].

\bibitem{FaucherGiguere:2010bq}
C.~A. Faucher-Giguere and A.~Loeb, \emph{{Pulsar-Black Hole Binaries in the
  Galactic Center}},
  \href{http://dx.doi.org/10.1111/j.1365-2966.2011.19019.x}{\emph{Mon. Not.
  Roy. Astron. Soc.} {\bf 415} (2011) 3951},
  [\href{http://arxiv.org/abs/1012.0573}{{\tt 1012.0573}}].

\bibitem{1982ApJ...253..908T}
J.~H. {Taylor} and J.~M. {Weisberg}, \emph{{A new test of general relativity -
  Gravitational radiation and the binary pulsar PSR 1913+16}},
  \href{http://dx.doi.org/10.1086/159690}{\emph{The Astrophys. J.} {\bf 253}
  (Feb., 1982) 908--920}.

\bibitem{S190814bv}
{Ligo Scientific Collaboration} and {VIRGO Collaboration}, \emph{{LIGO/Virgo
  S190814bv: Identification of a GW compact binary merger candidate.}},
  {\emph{GRB Coordinates Network} {\bf 25324} (Aug, 2019) 1}.

\bibitem{2019GCN.24168....1L}
{Ligo Scientific Collaboration} and {VIRGO Collaboration}, \emph{{LIGO/Virgo
  S190426c: Identification of a GW compact binary merger candidate.}},
  {\emph{GRB Coordinates Network} {\bf 24237} (Apr., 2019) 1}.

\bibitem{Lattimer:2019qdc}
J.~M. Lattimer, \emph{{The Properties of a Black Hole-Neutron Star Merger
  Candidate}},  \href{http://arxiv.org/abs/1908.03622}{{\tt 1908.03622}}.

\bibitem{AmaroSeoane:2012je}
P.~Amaro-Seoane et~al., \emph{{Low-frequency gravitational-wave science with
  eLISA/NGO}},
  \href{http://dx.doi.org/10.1088/0264-9381/29/12/124016}{\emph{Class. Quant.
  Grav.} {\bf 29} (2012) 124016}, [\href{http://arxiv.org/abs/1202.0839}{{\tt
  1202.0839}}].

\bibitem{Liu:2014uka}
K.~Liu, R.~P. Eatough, N.~Wex and M.~Kramer, \emph{{Pulsar-black hole binaries:
  prospects for new gravity tests with future radio telescopes}},
  \href{http://dx.doi.org/10.1093/mnras/stu1913}{\emph{Mon. Not. Roy. Astron.
  Soc.} {\bf 445} (2014) 3115--3132},
  [\href{http://arxiv.org/abs/1409.3882}{{\tt 1409.3882}}].

\bibitem{Cardoso:2015zqa}
V.~Cardoso, R.~Brito and J.~L. Rosa, \emph{{Superradiance in stars}},
  \href{http://dx.doi.org/10.1103/PhysRevD.91.124026}{\emph{Phys. Rev.} {\bf
  D91} (2015) 124026}, [\href{http://arxiv.org/abs/1505.05509}{{\tt
  1505.05509}}].

  
\bibitem{Day:2019bbh}
F.~V.~Day and J.~I.~McDonald,
\emph{{Axion superradiance in rotating neutron stars}},
 \href{http://dx.doi.org/10.1088/1475-7516/2019/10/051}
{\emph{JCAP} {\bf 10} (2019), 051}
[\href{http://arxiv.org/abs/1904.08341}{{\tt 1904.08341}}].

\bibitem{Kaplan:2019ako}
D.~E. Kaplan, S.~Rajendran and P.~Riggins, \emph{{Particle Probes with
  Superradiant Pulsars}},  \href{http://arxiv.org/abs/1908.10440}{{\tt
  1908.10440}}.

\bibitem{Dolan:2007mj}
S.~R. Dolan, \emph{{Instability of the massive Klein-Gordon field on the Kerr
  spacetime}}, \href{http://dx.doi.org/10.1103/PhysRevD.76.084001}{\emph{Phys.
  Rev.} {\bf D76} (2007) 084001}, [\href{http://arxiv.org/abs/0705.2880}{{\tt
  0705.2880}}].

\bibitem{East:2017ovw}
W.~E. East and F.~Pretorius, \emph{{Superradiant Instability and Backreaction
  of Massive Vector Fields around Kerr Black Holes}},
  \href{http://dx.doi.org/10.1103/PhysRevLett.119.041101}{\emph{Phys. Rev.
  Lett.} {\bf 119} (2017) 041101}, [\href{http://arxiv.org/abs/1704.04791}{{\tt
  1704.04791}}].

\bibitem{Pani:2012bp}
P.~Pani, V.~Cardoso, L.~Gualtieri, E.~Berti and A.~Ishibashi,
  \emph{{Perturbations of slowly rotating black holes: massive vector fields in
  the Kerr metric}},
  \href{http://dx.doi.org/10.1103/PhysRevD.86.104017}{\emph{Phys. Rev.} {\bf
  D86} (2012) 104017}, [\href{http://arxiv.org/abs/1209.0773}{{\tt
  1209.0773}}].
  
  
  \bibitem{Zhang:2019eid}
J.~Zhang and H.~Yang,
\emph{{Dynamic Signatures of Black Hole Binaries with Superradiant Clouds}},
 \href{http://dx.doi.org/10.1103/PhysRevD.101.043020}
  {\emph{Phys. Rev.} {\bf D101} (2020) 043020}, 
  \href{http://arxiv.org/abs/1907.13582}{{\tt 1907.13582}}.


  
  \bibitem{Wong:2019kru}
L.~K.~Wong,
\emph{{Superradiant Scattering by a Black Hole Binary}},
 \href{http://dx.doi.org/10.1103/PhysRevD.100.044051}{\emph{Phys. Rev.} {\bf
  D100} (2019) 044051},  \href{http://arxiv.org/abs/1905.08543}{{\tt 1905.08543}}.

\bibitem{Neilsen:2010ax}
D.~Neilsen, L.~Lehner, C.~Palenzuela, E.~W. Hirschmann, S.~L. Liebling, P.~M.
  Motl et~al., \emph{{Boosting jet power in black hole spacetimes}},
  \href{http://dx.doi.org/10.1073/pnas.1019618108}{\emph{Proc.Nat.Acad.Sci.}
  {\bf 108} (2011) 12641--12646}, [\href{http://arxiv.org/abs/1012.5661}{{\tt
  1012.5661}}].

  
  \bibitem{Poddar:2019zoe}
T.~Kumar Poddar, S.~Mohanty and S.~Jana,
  \emph{{Constraints on ultralight axions from compact binary systems}},
  \href{http://dx.doi.org/10.1103/PhysRevD.101.083007}{\emph{Phys. Rev.} {\bf
  D101} (2020) 083007}, [\href{http://arxiv.org/abs/1906.00666}{{\tt
  1906.00666}}].


\bibitem{Poddar:2019wvu}
  T.~Kumar Poddar, S.~Mohanty and S.~Jana,
   \emph{{Vector gauge boson radiation from compact binary systems in a gauged $L_\mu-L_\tau$ scenario}},
  \href{http://dx.doi.org/10.1103/PhysRevD.100.123023}{\emph{Phys. Rev.} {\bf
  D100} (2019) 123023}, [\href{http://arxiv.org/abs/1908.09732}{{\tt
  1908.09732}}].
  
  
  

\bibitem{Weisberg:2010zz}
J.~M. Weisberg, D.~J. Nice and J.~H. Taylor, \emph{{Timing Measurements of the
  Relativistic Binary Pulsar PSR B1913+16}},
  \href{http://dx.doi.org/10.1088/0004-637X/722/2/1030}{\emph{Astrophys. J.}
  {\bf 722} (2010) 1030--1034}, [\href{http://arxiv.org/abs/1011.0718}{{\tt
  1011.0718}}].

\bibitem{Simonetti:2010mk}
J.~H. Simonetti, M.~Kavic, D.~Minic, U.~Surani and V.~Vijayan, \emph{{A
  Precision Test for an Extra Spatial Dimension Using Black Hole--Pulsar
  Binaries}},
  \href{http://dx.doi.org/10.1088/2041-8205/737/2/L28}{\emph{Astrophys. J.}
  {\bf 737} (2011) L28}, [\href{http://arxiv.org/abs/1010.5245}{{\tt
  1010.5245}}].

\bibitem{Levin:2017mkq}
L.~Levin et~al., \emph{{Pulsar Searches with the SKA}},
  \href{http://dx.doi.org/10.1017/S1743921317009528}{\emph{IAU Symp.} {\bf 337}
  (2017) 171--174}, [\href{http://arxiv.org/abs/1712.01008}{{\tt 1712.01008}}].

\bibitem{Sesana:2016ljz}
A.~Sesana, \emph{{Prospects for Multiband Gravitational-Wave Astronomy after
  GW150914}},
  \href{http://dx.doi.org/10.1103/PhysRevLett.116.231102}{\emph{Phys. Rev.
  Lett.} {\bf 116} (2016) 231102}, [\href{http://arxiv.org/abs/1602.06951}{{\tt
  1602.06951}}].

\bibitem{Kuns:2019upi}
K.~A. Kuns, H.~Yu, Y.~Chen and R.~X. Adhikari, \emph{{Astrophysics and
  cosmology with a deci-hertz gravitational-wave detector: TianGO}},
  \href{http://arxiv.org/abs/1908.06004}{{\tt 1908.06004}}.

\bibitem{Sato:2017dkf}
S.~Sato et~al., \emph{{The status of DECIGO}},
  \href{http://dx.doi.org/10.1088/1742-6596/840/1/012010}{\emph{J. Phys. Conf.
  Ser.} {\bf 840} (2017) 012010}.

\bibitem{2007hsaa.book.....Z}
M.~{Zombeck}, \emph{{Handbook of Space Astronomy and Astrophysics: Third
  Edition}}.
\newblock 2007.

\bibitem{LIGOScientific:2018jsj}
{\scshape LIGO Scientific, Virgo} collaboration, B.~P. Abbott et~al.,
  \emph{{Binary Black Hole Population Properties Inferred from the First and
  Second Observing Runs of Advanced LIGO and Advanced Virgo}},
  \href{http://arxiv.org/abs/1811.12940}{{\tt 1811.12940}}.

\bibitem{Cardoso:2017kgn}
V.~Cardoso, P.~Pani and T.-T. Yu, \emph{{Superradiance in rotating stars and
  pulsar-timing constraints on dark photons}},
  \href{http://dx.doi.org/10.1103/PhysRevD.95.124056}{\emph{Phys. Rev.} {\bf
  D95} (2017) 124056}, [\href{http://arxiv.org/abs/1704.06151}{{\tt
  1704.06151}}].

\bibitem{Cook:1993qr}
G.~B. Cook, S.~L. Shapiro and S.~A. Teukolsky, \emph{{Rapidly rotating neutron
  stars in general relativity: Realistic equations of state}},
  \href{http://dx.doi.org/10.1086/173934}{\emph{Astrophys. J.} {\bf 424} (1994)
  823}.

\bibitem{Cipolletta:2015nga}
F.~Cipolletta, C.~Cherubini, S.~Filippi, J.~A. Rueda and R.~Ruffini,
  \emph{{Fast Rotating Neutron Stars with Realistic Nuclear Matter Equation of
  State}}, \href{http://dx.doi.org/10.1103/PhysRevD.92.023007}{\emph{Phys.
  Rev.} {\bf D92} (2015) 023007}, [\href{http://arxiv.org/abs/1506.05926}{{\tt
  1506.05926}}].

\bibitem{Hessels:2006ze}
J.~W.~T. Hessels, S.~M. Ransom, I.~H. Stairs, P.~C.~C. Freire, V.~M. Kaspi and
  F.~Camilo, \emph{{A radio pulsar spinning at 716-hz}},
  \href{http://dx.doi.org/10.1126/science.1123430}{\emph{Science} {\bf 311}
  (2006) 1901--1904}, [\href{http://arxiv.org/abs/astro-ph/0601337}{{\tt
  astro-ph/0601337}}].

\end{thebibliography}
\end{document}